\documentclass[epj]{webofc}
\usepackage[varg]{txfonts}   
\wocname{EPJ Web of Conferences}
\woctitle{ICNFP 2016}
\begin{document}
\selectlanguage{english}
\title{Deformations of spacetime and internal symmetries}
%
%

\author{Niels G. Gresnigt\inst{1,3}\fnsep\thanks{\email{niels.gresnigt@xjtlu.edu.cn}} \and
        Adam B. Gillard\inst{1} 
}

\institute{Department of Mathematical Sciences, Xi'an Jiaotong-Liverpool University, 111 Ren'ai Road, Suzhou SIP, Jiangsu, China 215123
}

\abstract{%
Algebraic deformations provide a systematic approach to generalizing the symmetries of a physical theory through the introduction of new fundamental constants. The applications of deformations of Lie algebras and Hopf algebras to both spacetime and internal symmetries are discussed. As a specific example we demonstrate how deforming the classical flavor group $SU(3)$ to the quantum group $SU_q(3)\equiv U_q(su(3))$ (a Hopf algebra) and taking into account electromagnetic mass splitting within isospin multiplets leads to new and exceptionally accurate baryon mass sum rules that agree perfectly with experimental data.

}
\maketitle
\section{Introduction}
\label{intro}

In this short article we discuss some of the applications of algebraic deformations to spacetime and internal symmetries. Deformation theory provides a systematic method of generalizing a physical theory through imposing additional fundamental constants. The passage from Galilean relativity to special relativity, with fundamental scale $c$, and from classical mechanics to quantum mechanics, with fundamental scale $\hbar$, can both be understood using deformations within the category of Lie algebras.

The combined Poincar\'e-Heisenberg Lie algebra structure that underlies the standard model (SM) may be deformed to a semi-simple algebra through the introduction of two additional fundamental length scales, at which point there no longer exist any non-trivial deformation within the category of Lie algebras and so no further fundamental scales can be introduced. 

The resulting algebra provides a natural candidate for the symmetry algebra in the low energy limit of the quantum-gravitational realm and an adaptation of physics to a framework that respects this algebraic structure provides a logical step toward an extension of the SM. Its energy momentum sector corresponds to the (anti) de Sitter algebra (a deformations of the Poincar\'e algebra). Among the possible new physics are the singleton representations, irreducible unitary representations on the anti de Sitter asymptotic boundary. 

A second class of deformations are $q$-deformations ($q$ for quantization). These are deformations of the universal enveloping algebra of a semi-simple Lie algebra into quantum groups. That is, these are deformations in the category of Hopf algebras. Both spacetime and internal symmetries may be $q$-deformed and quantum groups have found manifold applications in physics, including quantum gravity and gauge theories.

As a specific example of the application of $q$-deformations to internal symmetries, we consider a recent result by one of the authors that shows that replacing the flavor symmetry group $SU(3)$ by its quantum group counterpart $SU_q(3)$ and accounting the for mass splittings within baryon isospin multiplets leads to octet and decuplet baryon mass formulas of exceptional accuracy. We consider this as strong evidence for the conjecture that quantum groups play an important role in high energy physics.

\section{Lie-type deformations and the importance of stability}
Much of the formal theory relating to deformations of Lie algebras dates back to the works of Gerstenhaber \cite{gerstenhaber1964}, Nijenhuis and Richardson \cite{nijenhuis1967} in the 1960s. We provide here a brief overview only.

Given a Lie algebra $L_0$ with basis $\left\lbrace X_i \right\rbrace$ and bracket $\left[ X_i, X_j\right]_0 = if_{ij}^{k}X_k$, a one parameter deformation of $L$ is defined through the deformed commutator
\begin{eqnarray}
\left[ X_i, X_j\right]_t=\left[ X_i, X_j\right]_0+\sum\limits_{m=1}^{\infty}\phi_m(X_i,X_j)t^m,
\end{eqnarray}
where $t$ is the deformation parameter. Such a deformation is said to be trivial if $L_0$ is isomorphic to the deformed Lie algebra $L_t$. 

A trivial deformation is equivalent to saying there is an invertible linear transformation $T_t:V\rightarrow V$ such that
\begin{eqnarray}
T_t\left(\left[ X_i,X_j\right]_t\right)=\left[ T_tX_i, T_tX_j\right]_0.
\end{eqnarray} 
Under the linear transformations $X_i\rightarrow T_tX_i$, the structure constants transform as:
\begin{eqnarray}
f^{(t)\;k}_{ij}=T_i^uT_j^v(T^{-1})_w^k f_uv ^w,
\end{eqnarray}
and so the deformed commutators under a trivial deformation may be rewritten in terms of the $t$-dependent structure constants as
\begin{eqnarray}
\left[ X_i, X_j\right]_t=if^{(t)\;k}_{ij}X_k.
\end{eqnarray}
A Lie algebra is said to be stable (or rigid) if all infinitesimal deformations result in isomorphic algebras. In particular, semi-simple Lie algebras are stable.


This concept of stability provides insight into the robustness of a physical theory (the spacetime symmetries of which are encoded in a Lie algebra) or the need to generalize it. Given a Lie algebra describing symmetries of a physical theory, some of the structure constants depend on certain physical constants (for example $c$ for the Poincar\'e algebra and $\hbar$ for the Heisenberg algebra). The numerical values of these physical constants must be determined experimentally and are therefore not known without some error. A stable Lie algebra guarantees that the physics is not dependent on the precise value of these physical constants as any small deformation of the structure constants returns an isomorphic algebra. The same is not true for an algebra that is not stable. The concept of stability therefore provides insight into the validity of a physical theory or the need to generalize the theory. A theory with symmetries encoded within an unstable Lie algebra should be deformed until a stable Lie algebra is reached which encodes symmetries of wider validity and represents robust physics free of fine tuning issues. The importance of Lie-algebraic stability was first promoted by Mendes \cite{mendes1994}. Since then, several others have similarly argued that the stability of a physically relevant Lie algebra should be considered a physical principle \cite{chryssomalakos2004:1,ahluwaliakhalilova2005:1,gresnigt2007sph,ahluwalia2008ppa,gresnigt2015electroweak}.

\subsection{Lie-type deformations of spacetime symmetries}
Algebraically, much of the success of modern physics can be attributed to the Poincar\'e and Heisenberg algebras. It was noted by Faddeev (albeit in hindsight) that both the quantum and relativistic revolutions of the 20th century can be considered as Lie-algebraic stabilizations of the unstable algebras of classical mechanics and Galilean relativity respectively \cite{faddeev1989}. The stabilization of these algebras give the Heisenberg and Poincar\'e algebras, which are both individually stable\footnote{Technically it is the semi-simple Lorentz algebra that is stable. Adding the abelian translations renders the Poincar\'e algebra unstable again. Its stabilized version is the (anti) de Sitter algebra. However, the instability of the Poincar\'e algebra may be ignored as long as one restricts oneself to the tangent space.}. The deformation parameters introduced in the process of deforming an algebra correspond to new invariant scales, $c$ and $\hbar$ in the examples above. Lie algebra deformations therefore provide a systematic way of introducing new invariant scales.

It has been argued in \cite{ahluwalia2008ppa} that in quantum cosmology/gravity, an operationally defined view of physical space will inevitably ask for both a minimal (Planck order) length scale as well as a cosmological length scale, and furthermore, that algebraically there should be a mechanism that describes how the Poincar\'e -algebraic description of present day spacetime relates to the conformal algebraic description of spacetime in an early universe where quarks and leptons had yet to acquire mass. 

Mathematically, a Lie algebra that incorporates these requirements already exists. Although the Poincar\'e and Heisenberg algebras are separately stable, the combined Poincar\'e-Heisenberg algebra, consisting of the Poincar\'e algebra extended by the position generators and their commutation relations with momenta, lacks the desired stability. Its stabilized form was found by \cite{mendes1994}. The resulting stabilized Poincar\'e-Heisenberg algebra (SPHA) turned out to be (up to various signs) the same algebra arrived at by Yang \cite{yang1947} in 1947 based on the work of Snyder who demonstrated that the assumption that space be a continuum is not required for Lorentz invariance earlier the same year \cite{snyder1947}. Uniqueness, an issue that was not addressed by Mendes, was later demonstrated by Chryssomalakos and Okon \cite{chryssomalakos2004:1}. SPHA introduces in addition to $c$ and $\hbar$, two invariant length scales $\ell_P$ and $\ell_C$, the former on the order of the Planck length, the latter a cosmological length scale. Furthermore, it was shown in \cite{ahluwalia2008ppa} that in a certain limit, this algebra satisfies the commutation relations of the conformal algebra of spacetime. In the contraction $\ell_P\rightarrow 0$, the algebra reduces to the (anti) de Sitter algebra; the deformation of the Poincar\'e algebra.

\subsection{Singleton physics}

The Poincar\'e symmetries of Minkowski spacetime are deformed to those of (anti) de Sitter spacetime through the introduction of the length scale $\ell_C=\Lambda^{-\frac{1}{2}}$. This introduces interesting new features. Elementary particles in Minkowski spacetime are associated with unitary irreducible representations of the Poincar\'e group (the isometry group of Minkowski spacetime).  In the present case, elementary particles should likewise be associated with the unitary irreducible representations not of the Poincar\'e group, but of the (anti) de Sitter group. 

The representations of the anti de Sitter group, like the Poincar\'e group, admit a positive minimum energy. This means that representations of this group naturally lends themselves to a particle interpretation. This is not the case for the de Sitter group. The most fundamental irreducible representations of the anti de Sitter group were first discovered by Dirac \cite{Dirac1963}, and are called the singleton representations. In anti de Sitter space, massless particles are composed of two singletons \cite{flato1978one}. These singletons are naturally confined (in a kinematic sense) which has also led to the question of whether perhaps singletons take the role of quarks \cite{flato1986}. Compatibility between the singleton representations and quantum electrodynamics (QED) was demonstrated by in \cite{flato1981quantum}.  

The two singleton representations, named \textit{Di} and \textit{Rac} are given by
\begin{eqnarray}
Rac=D(\frac{1}{2}, 0),\qquad Di=D(1,\frac{1}{2}).
\end{eqnarray}
These singleton representations have the interesting property that a direct product of two positive energy singletons reduces to a sum of massless representations of $SO(2,3)$ as follows ($s$ is the spin) \cite{flato1978one}
\begin{eqnarray}
Rac\otimes Rac&=&\oplus_{s=0,1,...}D(s+1,s),\\
Rac\otimes Di&=&\oplus_{2s=1,3,...}D(s+1,s),\\
Di\otimes Di&=&\oplus_{s=1,2,...}D(s+1,s)\oplus D(2,0).
\end{eqnarray}

The $Di$ and $Rac$ themselves do not have contractions to representations of the Poincar\'e group. In the flat space limit the singletons reduce to vacua. It may then be possible to consider all leptons as \textit{Rac}-\textit{Di} and the vector mesons of the electroweak model as \textit{Rac}-\textit{Rac} composites \cite{Sternheimer2007}.

\section{q-deformations}
\label{sec-1}
A stable Lie algebra cannot be deformed any further within the category of Lie algebras. It is however possible to deform the completed universal enveloping algebra of a Lie algebra. Such a $q$-deformation is a deformation in the category of Hopf algebras and deforms the universal enveloping algebra into a quantum group. 

Quantum groups (which are algebras rather than groups) provide a generalization of familiar symmetry concepts through extending the domain of classical group theory. Quantum groups depend on a deformation parameter $q$ with the value $q=1$ returning the undeformed universal enveloping algebra but for general $q$ giving the structure of a Hopf algebra. First formalized by Jimbo \cite{jimbo1985aq} and Drinfeld \cite{drinfeld1985soviet} as a class of Hopf algebras, quantum groups have found many applications in theoretical physics, see \cite{Finkelstein2012,Majid1994,Lukierski2003,castellani1996quantum} and references therein.

\subsection{The quantum groups $SU_q(n)\equiv U_q(su(n))$}

The literature on quantum groups and algebras is extensive. For an excellent introduction the reader is directed to \cite{jaganathan2001some}.

The quantum (enveloping) algebra $SU_q(n)\equiv U_q(su(n))$ corresponding to a one-parameter deformation of the universal enveloping algebra of $su(n)$, is a Hopf algebra with unit $\mathbf{1}$ and generators $H_i$, $X_i^{\pm}$, $i=1,2,...,n-1$, defined through the commutation relations in the Cartan-Chevalley basis as 
\begin{eqnarray}
\left[ H_i, H_j\right] &=&0\\
\left[ H_i, X_j^{\pm}\right] &=&a_{ij}X_j^{\pm}\\
\left[ X_i^+, X_j^-\right]&=&\delta_{ij}[H_i]_q\equiv\delta_{ij}\frac{q^{H_i}-q^{-H_i}}{q-q^{-1}}\label{quantalg3},
\end{eqnarray}
together with the quadratic and cubic deformed $q$-Serre relations
\begin{eqnarray}
\left[ X_i^{\pm}, X_j^{\pm}\right]=0,\;\; j\neq i\pm 1,\;\; 1\leq i,j\leq n-1
\end{eqnarray}
and
\begin{eqnarray}\label{cubicserre}
(X_i^{\pm})^2 X_j^{\pm}-[2]_q X_i^{\pm}X_j^{\pm}X_i^{\pm}+X_j^{\pm}(X_i^{\pm})^2=0,\;\; j=i\pm 1,\;\; 1\leq i,j\leq n-1
\end{eqnarray}
respectively \cite{jimbo1985aq,quesne1992complementarity}. Here $a_{ij}$ is an element of the Cartan matrix
 \begin{displaymath}
   a_{ij} = \left\{
     \begin{array}{lr}
       2 & j=i\\
       -1 & j=i\pm 1\\
       0 & \textrm{otherwise}.
     \end{array}
   \right.
\end{displaymath} 
The $q$-number
\begin{eqnarray}
[N]_q=\frac{q^N-q^{-N}}{q-q^{-1}}
\end{eqnarray}
is defined for both operators (as in equation (\ref{quantalg3})) and real numbers\footnote{In this paper we will only have to deal with integer values of $N$. The definition however holds for real numbers.} (as in equation (\ref{cubicserre})). The definition of the algebra is completed by the Hermiticity properties
\begin{eqnarray}
(H_i)^{\dagger}=H_i,\qquad (X_i^{\pm})^{\dagger}=X_i^{\mp}.
\end{eqnarray}

The quantum algebra $SU_q(n)$ has the structure of a Hopf algebra admitting a coproduct, counit and antipode. These are not used here and so we do not define them (see \cite{quesne1992complementarity}). In the limit $q=1$ the above relations approach the relations for the universal enveloping algebra $U(su(n))$ but for general $q$ they represent a deformation of the universal enveloping algebra of $su(n)$.

\subsection{$q$-deformations of spacetime symmetries}

Quantum groups as well as non-commutative spaces have been utilized in many approaches to quantum gravity. There is strong evidence that in both 2+1 and 3+1 quantum gravity, the symmetry algebra must be $q$-deformed. Quantum gravity often requires a nonzero cosmological constant to be well defined. In loop quantum gravity (LQG) for example, the only known exact state in the connection basis that has semi-classical properties, the Kodama state, exists only when $\Lambda$ is nonzero \cite{smolin2002quantum}. 

In the quantization of general relativity, the local gauge symmetry of the spacetime connection must be $q$-deformed from $SU(2)$ to $SU_q(2)$ \cite{PhysRevD.70.065020}. As explained in \cite{bianchi2011note}, for a universe characterized by a maximal $\ell_C=\Lambda^{\frac{1}{2}}$ and minimal length $\ell_P$ the local rotational symmetry is better described by the quantum groups $SU_q(2)$ than by $SU(2)$. In this case the deformation parameter $q$ given by
\begin{eqnarray}
q=\exp{(i\ell_P^2\Lambda)}.
\end{eqnarray}
It is interesting that it is precisely these two scales that are obtained in the deformation of the Poincar\'e -Heisenberg algebra. It seems therefore that both Lie deformations and $q$-deformations play an important role in quantum spacetime.

Quantum spacetime is generally accepted to be non-commutative. The simplest example of a non-commutative geometry is provided by the quantum plane. One may quantize the classical vector space of commuting coordinates $x$ and $y$ to obtain a quantum vector space by assuming that the coordinates no longer commute with each other. This noncommutativity of the quantized coordinates $X$ and $Y$ of the quantum plane can be modelled by
\begin{eqnarray}
XY=qYX,
\end{eqnarray}
where $q$ is the deformation parameter. In the classical limit $q\rightarrow 1$, one recovers the classical plane. As in the classical case, it turns out to be possible to define a differential calculus on the quantum plane that is invariant under a generalization of the classical group $SL(2)$. This generalization of the classical group is the quantum group $SL_q(2)$. Thus, the theory of quantum groups is part of the program of noncommutative geometry.

\subsection{$q$-deformed electroweak theory}

Finkelstein in a series of papers, see \cite{Finkelstein2010} and references therein, developed a solitonic model of elementary particles in terms of quantized knots. The model is constructed by replacing the electroweak gauge group $SU(2)\times U(1)$ by the quantum group $SUq(2)$. Quantum groups are closely related to the Jones polynomial of knots and $SU_q(2)$ is the algebra of oriented knots. It is this close relation between quantum groups and knots that allows for a deformed gauge theory based on $SU_q(2)$ in which the quantum numbers of solitonic elementary particles are topologically defined as certain knot invariants.

Assuming that the most elementary particles are also the most elementary knots, Finkelstein identified the four classes of fermions in the SM are identified with the four quantum trefoils. The electroweak gauge bosons are represented as compositions (knot sums) of trefoil knots. The quantized knots are labelled by the irreducible representations of $SU_q(2)$ with the elementary fermions described by the trefoil representations of $SU_q(2)$. The three femions of each family (for example electron, muon, and tau) correspond to the three lowest states in the excitation spectrum of the knots. The model successfully matches knots with the electroweak field quanta. 

The model was later expanded into a preon model in which the knotted field has a composite structure of three or more preons that are described by the fundamental representation of $SU_q(2)$. Interestingly, the preons in this model correspond to twisted loops rather than non-trivial knots. Given the close relation between knots and braids (knots being the closure of a braid) these ideas also seem to be closely related to another preon inspired model, called the Helon model, in which the simplest non-trivial braids consisting of three ribbons and two crossings map precisely to the first generation of SM leptons and quarks \cite{Bilson-Thompson2005}. It was later demonstrated that these braids may be embedded within spin networks, which makes it compatible with loop quantum gravity in the spin network basis \cite{Bilson-Thompson2007,Bilson-Thompson2012}.

\subsection{$SU_q(3)$ flavor symmetry}
As a more detailed application of $q$-deformations to particle physics, we review the recent result by one of us, see \cite{gresnigt2016charge}, where it is shown that replacing the flavor symmetry group $SU_q(3)$ by the quantum group $SU_q(3)$ and accounting for the electromagnetic mass splitting of baryons within isospin multiplets leads to exceptionally accurate baryon mass formulas. 

The standard Gell-Mann-Okubo mass formula \cite{gell1961eightfold,okubo1962note} is
\begin{eqnarray}
M=\alpha_0+\alpha_1 S+\alpha_2 \left[ I(I+1)-\frac{1}{4} S^2\right], 
\end{eqnarray}
where $M$ is the mass of a hadron within a specific multiplet, $S$ and $I$ are the strangeness and isospin respectively, and $\alpha_0, \alpha_1, \alpha_2$ are free parameters. By eliminating the free parameters $\alpha_0, \alpha_1, \alpha_2$, one obtains mass relations between the different baryons within a given (isospin) multiplet. 
For the case of octet baryons one obtains the standard relation
\begin{eqnarray}\label{GMOoctet}
N+\Xi=\frac{3}{2}\Lambda+\frac{1}{2}\Sigma,
\end{eqnarray} 
whereas for the decuplet baryons $+\frac{3}{2}$ one obtains the equal spacing rule
\begin{eqnarray}\label{GMOdecuplet}
\Delta-\Sigma^*=\Sigma^*-\Xi^*=\Xi^*-\Omega
\end{eqnarray}
These formulas hold to first order in flavor symmetry breaking only. Using the most recent data formula (\ref{GMOoctet}) is accurate to about 0.6\%. The equal spacing rule (\ref{GMOdecuplet}) is less accurate however a modified relation $\Omega-\Delta=3(\Xi^*-\Sigma^*)$ due to Okubo \cite{okubo1963some} is accurate to about 1.4\%.

The idea of replacing the classical flavor symmetry group by its quantum group counterpart was first considered in \cite{gavrilik1997quantum,gavrilik2001quantum,gavrilik2004quantum}, and its authors found meson and baryon mass sum rules of improved accuracy. By fixing a definite value for the deformation parameter $q$ (through fitting the data), the modified baryon mass sum rules are accurate to an impressive 0.06\% for the octet baryons and 0.32\% for the decuplet baryons.

The basic approach of the construction is the representation theory of $U_q(su(n))$ \cite{gavrilik1995representations,gavrilik2004quantum}. $q$-Deformed mass sum rules are computed from the expectation value of the mass operator which is defined in terms of the generators of the dynamical algebras (quantum groups) $U_q(u(n+1))$ or $U_q(u(n,1))$. The expectation values are computed from the matrix elements of these generators. Utilizing the $q$-algebras $U_q(u(n+1))$ or $U_q(u(n,1))$ of dynamical symmetry, breaking of $n$-flavor symmetries up to exact isospin symmetry $SU_q(2)$ are realized and the $q$-analogues of mass sum rules for baryon multiplets are derived. 

The masses used in both the standard as well as the $q$-deformed mass formulas are the averages of the isospin multiplets (isoplets). At the level of accuracy of the $q$-deformed mass relations, the mass splittings within isoplets become significant and can no longer be ignored. For example $\Xi^{*-}-\Xi^{*0}=3.2$MeV and $\Sigma^{*-}-\Sigma^{*+}=4.4$MeV, representing about $\sim 0.2-0.3\%$ of the average isoplet mass. The impressive accuracy of the deformed mass formulas lose their significance when these mass splittings (due to electromagnetic contributions to the masses of baryons) are not considered. 

The electromagnetic contributions to baryon masses are determined within the QCD general parametrization scheme in the spin-flavour space considered by Morpurgo \cite{morpurgo1989field} and to zeroth order symmetry breaking the electromagnetic contributions to the octet baryon masses are given in terms of four parameters \cite{morpurgo1992new}. 



By replacing the classical flavor group with its associated quantum group as well as accounting for the electromagnetic mass splittings within isospin multiplets, charge specific baryon mass formulas with $SU_q(3)$ flavor symmetry were found by one of us \cite{gresnigt2016charge}. For the octet baryons the new mass formula is
\begin{eqnarray}
p+\frac{\Xi^{0}}{[2]_{q}-1}=\frac{\Lambda^0}{[2]_{q}-1}+(2\Sigma^+-\Sigma^0),
\end{eqnarray}
where $q=e^{i\pi/n}$ for integer $n$, is the deformation parameter. Substituting experimental data \cite{olive2014review} we find that the closest fit to the data is when $n=7$ so that $q_7=e^{i\pi/7}$, in agreements with \cite{gavrilik1997quantum}. The formula is
\begin{eqnarray}\label{newoctet}
p+\frac{\Xi^{0}}{[2]_{q_7}-1}=\frac{\Lambda^0}{[2]_{q_7}-1}+(2\Sigma^+-\Sigma^0).
\end{eqnarray}
The formula gives LHS=2577.87MeV and RHS=2577.33MeV, which has an error of only 0.02\%. This is a significant improvement on the standard octet relations (\ref{GMOoctet}) which holds to 0.57\% as well as the $q$-deformed relation \cite{gavrilik2001quantum} which (for $q=e^{i\pi/7}$) holds to 0.06\%.

For the decuplet baryons, the mass formula is
\begin{eqnarray}\label{newdecuplet}
\Omega^--\Delta^-=([2]_q+1)(\Xi^{*-}-\Sigma^{*-}),
\end{eqnarray}
where $q$ is again of the form $q=e^{i\pi/n}$ for integer $n$. In \cite{gavrilik1997quantum} is was suggested that choosing $q=e^{i\pi/14}$ provides a good fit to data. Although this remains true in the present case, there are other values of $q$ for which the error is smaller. In particular, we consider the case where $q=e^{i\pi/21}$ \footnote{Although this is not the absolute best choice for $q$, because 21 is a multiple of 7, it allows for an elegant new relation between octet and decuplet baryons \cite{gresnigt2016charge} (this is also the reason why $n=14$ is chosen in \cite{gavrilik1997quantum,gavrilik2001quantum}).}.

The new baryon mass formulas based on an $SU_q(3)$ flavor symmetry are accurate to well within the experimental uncertainty. It may be that nature's true flavor symmetry is not $SU(3)$ but rather $SU_q(3)$. 

\section{Conclusion}
The deformations of spacetime and internal symmetries were considered. For deformation within the Lie algebra framework, the concept of stability provides a important role in generalizing the symmetries of a physical theory in a systematic way. An unstable symmetry algebra should be deformed, via the introduction of additional invariant scales, into a stable one. The resulting theory will describe robust physics and be free of fine tuning issues. 

Deformations within the category of Lie algebras are not the only type of deformations that have been considered in the literature. For example, in double special relativity (DSR) the deformations considered introduce non-linearity into the theory (i.e. the Lie-algebraic framework is abandoned) \cite{Amelino-Camelia2002b,Kowalski-Glikman2002a,girelli2005deformed}. The advantage of Lie-type deformations is that they can be handled systematically without introducing non-linearity. Indeed, it has been shown that through a suitable redefinition of the generators, triply special relativity \cite{kowalskiglikman2004tsr} can be brought into linear form \cite{chryssomalakos2004:2}.

Deforming the combined Poincar\'e plus Heisenberg algebra introduces two new invariant length scales. The energy-momentum sector of the stable algebra is the (anti) de Sitter algebra. For the case of the anti de Sitter spacetime this leads to interesting new physics, where particles correspond to the unitary irreducible representations of not the Poincar\'e group but rather the anti de Sitter group $SO(2,3)$. The singleton representations of this group suggest that massless particles may be considered as singleton composites.

Once a stable algebra is obtained, no further non-trivial deformations are possible within the framework of Lie algebras. One can still however deform in the category of Hopf algebras. This gives rise to quantum groups. There is evidence from quantum gravity that spacetime symmetries should be $q$-deformed and in this case the deformation parameter can be expressed in terms of the two length scales introduced in the deformation of the Poincar\'e -Heisenberg algebra.

Quantum groups also likely play a role in generalizing the symmetries of SM particles. The use of quantum groups in gauge theories allows for a solitonic description of elementary particles where a particle's quantum numbers are described in terms of knot invariants. Taken as a flavor symmetry, the quantum group $SU_q(3)$ gives rise to exceptionally accurate baryon mass formulas that agree perfectly with the experimental data. 

%
%
\bibliography{NielsReferences}  


%
%

\end{document}